\documentclass[10pt,twocolumn,letterpaper]{article}

\usepackage{ijcb}

\usepackage{times}
\usepackage{epsfig}
\usepackage{graphicx}
\usepackage{amsmath}
\usepackage{amssymb}
\usepackage{algpseudocode}
\usepackage{algorithm}
\usepackage{algcompatible}
\usepackage{multicol}
\usepackage{multirow}
\usepackage{makecell}
\usepackage{hyperref}
\usepackage{bbold}

\ijcbfinalcopy 

\ifijcbfinal\fi
\begin{document}

\title{Straight Through Gumbel Softmax Estimator based Bimodal Neural Architecture Search for Audio-Visual Deepfake Detection}
\author{
Aravinda Reddy PN$^{1}$\thanks{Equal contribution.} \and
Raghavendra Ramachandra$^{2}$ \and
Krothapalli Sreenivasa Rao$^{2,1}$ \and
Pabitra Mitra$^{3,1}$ \and
Vinod Rathod$^{1}$ \thanks{Equal contribution.} \and
\vspace{0.2cm}
Indian Institute of Technology Kharagpur$^{1}$ Norwegian University of Science and Technology$^{2}$
}


\maketitle
\thispagestyle{empty}

\begin{abstract}
Deepfakes are a major security risk for biometric authentication. This technology creates realistic fake videos that can impersonate real people, fooling systems that rely on facial features and  voice patterns for identification. Existing multimodal deepfake detectors rely on conventional fusion methods, such as majority rule and ensemble voting, which often struggle to adapt to changing data characteristics and complex patterns. In this paper, we introduce the Straight-through Gumbel-Softmax (STGS) framework, offering a comprehensive approach to search multimodal fusion model architectures. Using a two-level search approach, the framework optimizes the network architecture, parameters, and performance. Initially, crucial features were efficiently identified from backbone networks, whereas within the cell structure, a weighted fusion operation integrated information from various sources. An architecture that maximizes the classification performance is derived by varying parameters such as temperature and sampling time. The experimental results on the FakeAVCeleb and SWAN-DF datasets demonstrated an impressive AUC value 94.4\% achieved with minimal model parameters.

\end{abstract}

\section{Introduction}
While deep generative models have led to incredibly realistic synthetic audio and video \cite{korshunov2018deepfakes}, this technological leap presents a major security challenge. These advancements can be exploited to bypass biometric authentication systems, which rely on a person's unique physical or vocal characteristics. For example, visual deepfakes utilize facial manipulation techniques to alter identities, depict malicious actions, and manipulate facial attributes. Additionally, recent progress in deepfake technology has facilitated real-time cloning of human voices \cite{chintha2020recurrent}. Human voice cloning methods utilize neural networks to synthesize speech samples that closely resemble the target speaker's voice, posing further challenges for authentication systems and enabling the impersonation of individuals such as celebrities and politicians, as well as financial fraud. Accordingly, various unimodal detectors for audio and visual deepfakes have been proposed \cite{chintha2020recurrent}. Current studies \cite{mittal2020emotions, chugh2020not, hashmi2022multimodal, feng2023self} use the majority rule, ensemble voting, and fusion module to fuse audio and visual features for accurate audio-visual deepfake detection; however, these methods are unable to adapt to changing data characteristics, limited capacity to learn complex patterns, and relationships \cite{pham2018neural}.

Recently, with advancements in Neural Architecture Search (NAS), NAS has shown great potential for use in multimodal learning and the fusion of each modality. Recently, deep multimodal Neural Architecture Search (MMNAS) \cite{yu2020deep} has permitted attention operations to be searched; however, during architecture search, the network's topological structure is fixed. In another study \cite{yin2022bm}, multimodal DNNs were searched for using both fusion and unimodal feature selection strategies using a simple Softmax. In another study, \cite{pang2022gumbel, chang2019differentiable} utilized the Gumbel-Softmax distribution for NAS and analyzed the effect of sampling times for medical image analysis and image classification via gradient estimation; however, they did not consider the effect of temperature parameters that control the entropy of distribution.
\begin{figure}[!h]
  \centering
   \includegraphics[width=1\linewidth]{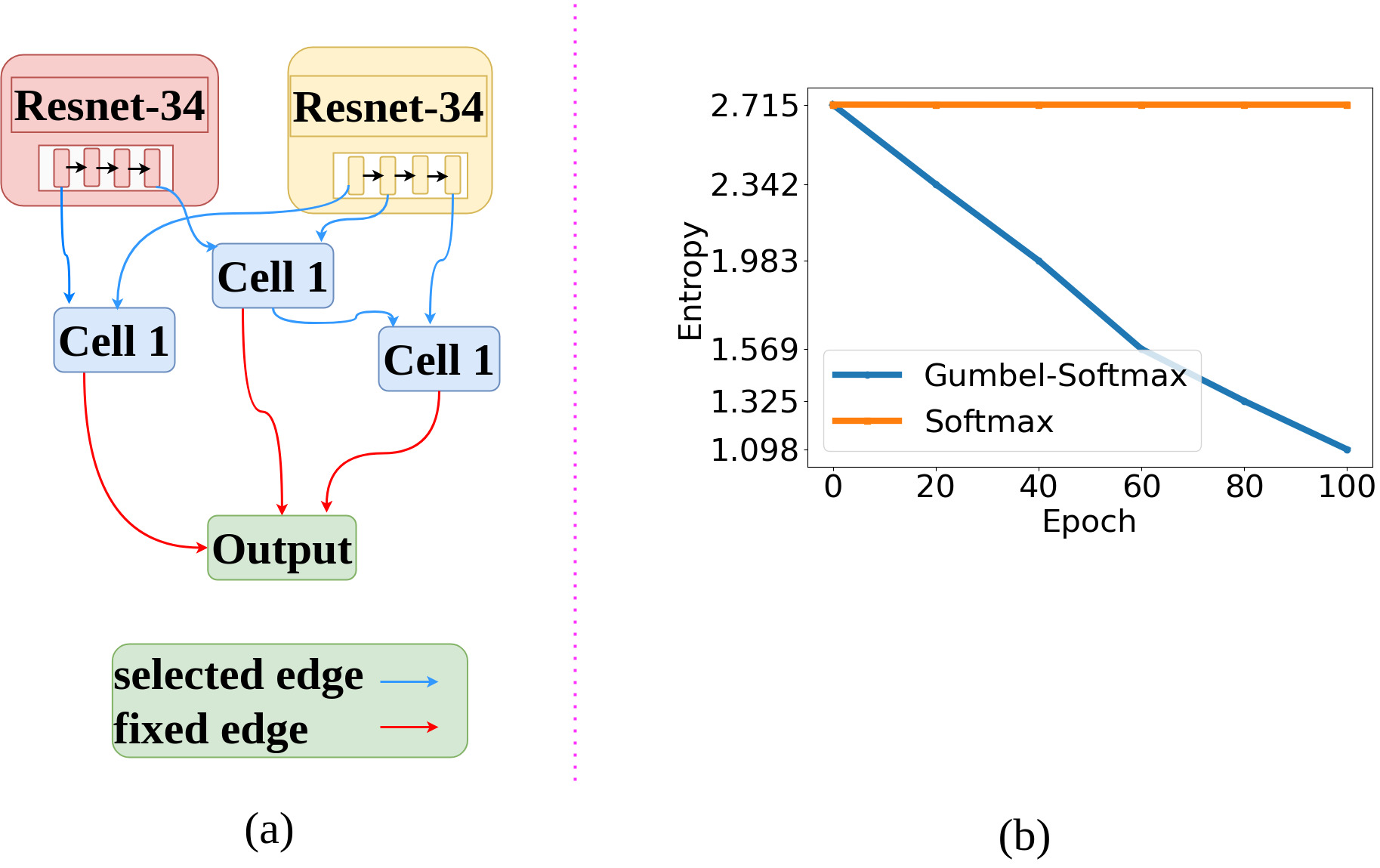}

   \caption{An overview of our proposed STGS-BMNAS for AV Deepfake detection. (a) Two level search based architecture b) Average entropy plot for two learnable parameters for the proposed STGS-BMNAS and Softmax \cite{yin2022bm}.}
   \label{fig:mini_arch}
\end{figure}

The aim of this work is to develop an automatic architecture for reliable audio-video fake detection. To this extent, we propose a novel method called Straight through Gumbel-Softmax estimator-based Bi-modal Neural architecture search (STGS-BMNAS) to adaptively learn the architectures of DNNs for audio-visual deepfake detection. STGS-BMNAS adopts a two-level search scheme in which it learns the unimodal feature selection from the first level by sampling the search space using Gumbel Softmax. In the second level, we utilize a multimodal weighted fusion strategy and vary the temperature parameter along with sampling from the search space via Gumbel Softmax. The sampling is performed to increase the search space for the primitive operations, so that maximum of the softmax of the primitive operation is picked up from the pool of operations. STGS-BMNAS utilises Straight-Through Estimator to handle the non-differentiability introduced by the Gumbel-Softmax trick during backpropagation. It replaces the gradient computation with a simpler operation to pass the gradients through the sampling process. Figure \ref{fig:mini_arch} shows the proposed STGS-BMNAS, where the first level consists of a backbone network from which the unimodal features are drawn and fused into the cells. A cell is a  directed acyclic graph (DAG) consisting of ordered sequences on a node. Each node is a latent representation and has directed edges associated with operations (primitive operations) that transform the node. We trained our proposed STGS-BMNAS and conventional Softmax \cite{yin2022bm} and plotted the average entropy plot from a two-level search. We show that the entropy value decreases compared to regular softmax, as shown in Figure \ref{fig:mini_arch} and conclude that the architecture converges when compared to regular softmax. We conducted experiments on the proposed STGS-BMNAS in an end-to-end framework. STGS-BMNAS shows superior performance compared with state-of-the-art audio-visual deepfake detection. Our proposed framework achieved comparable performance with fewer GPU days and fewer model parameters. To the best of our knowledge, STGS-BMNAS is the first bi-modal NAS framework to derive an optimal architecture for audio-visual deepfake detection.
The main contributions of this paper are as follows:

\begin{enumerate}
    \item To achieve a more generalised and robust design of DNNs in multimodal learning, we propose a new approach that uses Straight Through Gumbel-softmax based NAS to search for both unimodal feature selection strategy and weighted multimodal fusion strategy.
    \item We present a novel STGS-BMNAS framework to detect the audio-visual deepfake detection. STGS-BMNAS is an end-to-end multimodal fusion model that is fully searchable using a two-level schema. During back propagation, the gradients are passed using a straight-through estimator. The source code of the proposed method will be released after acceptance of the paper.
    \item We evaluate the proposed STGS-BMNAS by conducting extensive experiments for audio-visual deepfake detection. Empirical values suggest that STGS-BMNAS produces fewer modal parameters with comparable performance when compared to previous state-of-the-art methods.
\end{enumerate}

\section{Related work}
Advanced deep learning techniques have been employed to manipulate and synthesize both visual and auditory aspects, yielding highly realistic fake videos.
Some studies have aimed to identify the inconsistencies between audio and video content. For instance, the approach outlined in \cite{korshunov2019tampered} capitalizes on the limitations of certain generation methods to accurately synchronize audio streams with video content. Similarly, \cite{zhou2021joint} focused on leveraging the inherent synchronization between video and audio. However, with rapid technological advancements, numerous methods, such as those discussed in \cite{prajwal2020lip}, can now generate highly realistic deepfakes with accurately synchronized speech and lip movements, posing significant challenges for audiovisual synchronization analysis. The approach presented in \cite{mittal2020emotions} concentrates on extracting emotion features from both modalities and conducting similarity analysis within the same audio and video. In \cite{wang2022m2tr}, a multimodal and multiscale transformer was crafted to leverage spatial and frequency domain artifacts. Meanwhile, \cite{chugh2020not} sought inconsistencies between audio and visual streams by training a modality dissonance score. In \cite{cheng2023voice}, an efficient multimodal matching framework was developed to distinguish real and fake videos. Building on this matching approach, we improve the traditional contrastive loss to align the objectives between the generic and deepfakes. A new self-supervised approach, where a video forensic method is rooted in anomaly detection, is capable of identifying such inconsistencies using solely real, unlabeled data for training. Multimodal trace extracts \cite{raza2023multimodaltrace} learn channels from audio and visual modalities, independently mixing them in the IntrAmodality Mixer Layer (IAML). Subsequently, they were jointly processed in IntErModality Mixer Layers (IEML) before being fed to a multilabel classification head.

\section{Proposed Method: STGS-BMNAS}
\label{sec:method}
This work presents a novel framework called STGS-BMNAS for multimodal feature fusion and operation exploration through Neural Architecture Search (NAS). At the first level, features from the backbones and cells were explored within the DAG. This DAG outputs the architecture weights, indicating the importance of the operation at each edge. The second level involves a DAG of the nodes within a cell, each representing an operation chosen from a predefined pool. In the following subsections, we discuss the proposed STGS-BMNAS framework for AV fake detection. 
\subsection{Gumbel distribution}
The Gumbel distribution \cite{gumbel1935valeurs} is a specific case (Type I) within the broader family of generalized extreme value distribution, designed to characterize extremes and rare occurrences. A random variable following the Gumbel distribution, frequently denoted as a 'Gumbel' in this context,  is defined by two parameters: location parameter $\mu \in \mathbb{R}$ and non-negative scale parameter $\beta \in \mathbb{R}_{\geq 0}$. The corresponding probability density and cumulative density functions are given by:

\begin{equation}
    f(x)=\frac{1}{\beta}e^-{\frac{x-\mu}{\beta}}e^{-e^{-\frac{x-\mu}{\beta}}}
    \label{eq:pdf}
\end{equation}

\begin{equation}
    F(x)=e^{-e^{-\frac{x-\mu}{\beta}}}
    \label{eq:cdf}
\end{equation}

The inverse cumulative density function (ICDF) is also called quantile function given by:

\begin{equation}
    F^{-1}(u)=-\beta log(-log(u))+\mu
    \label{eq:icdf}
\end{equation}
Equation \ref{eq:icdf} is used in inverse transform sampling to transform sample from Uniform distribution $U(0,1)$ into a Gumbel via a double logarithmic relation. 

\subsection{Gumbel-max trick}
The Gumbel-max trick is a method for sampling from a categorical random variable $I\sim Cat(\pi)$. This technique entails adding independent and identically distributed Gumbel noise samples to the un-normalized log probabilities and subsequently choosing the index with the highest value. This chosen index follows a Gumbel distribution.  More specifically:
\begin{equation}
    I=\mathop{argmax}_{i \in D}\:\{G^{(i)}+log \:\theta_i\}\sim Cat(\pi)
\end{equation}
where $G^{(1)}, G^{(2)}, G^{(3)}... G^{(D)}$ are the i.i.d samples drawn from Gumbel distribution \ref{eq:icdf}.
\subsection{Gumbel-Softmax distribution}
Rather than generating discrete/hard samples from an (unstructured) categorical distribution, it is possible to define soft samples, particularly useful for gradient estimation. To understand the connection between these hard and soft samples, it is necessary to examine the hard samples represented in their one-hot embedding format i,e., $\mathbb{1}_\omega \in \{0,1\}^N$. 
\begin{equation}
    z=onehot(\mathop{argmax}_{i \in D}\:\{G^{(i)}+log \:\theta_i\})
\end{equation}

A soft sample $S_\lambda$ can be characterized as a vector of equal length, where the distribution mass is distributed across multiple bins instead of being concentrated in a single class. Concurrent works \cite{jang2016categorical,maddison2016concrete} have introduced the Gumbel-Softmax or Concrete distribution, of which an exact sample is a relaxation of $\mathbb{1}_\omega$. From \cite{jang2016categorical, maddison2016concrete} we derive PDF of this distribution and denoted by $GS(\pi,\lambda).$ More specifically, the $i^{th}$ index of soft sample $S_\lambda \: \in \{\mathbb{R}^N_{\geq 0}:|S_\lambda|=1\}$ is defined as: 

\begin{equation}
    S_{i;\lambda}=\frac{exp((log \: \theta_i+\: G^{(i)})/\lambda)}{\sum_{j \: \in D}exp((log \: \theta_i)+G^{(j)})/\lambda)}
    \label{eq:act_gumbel}
\end{equation}

Here, $\lambda$ represents a temperature parameter impacting the entropy of both the Gumbel-Softmax distribution and the associated samples. When considering these samples as relaxations of samples from $Cat(\pi)$, $\lambda$ can be interpreted as a parameter defining the degree of relaxation for the soft sample $S_\lambda$. As the softmax temperature $\lambda$ approaches 0, samples generated from the Gumbel-Softmax distribution transition to a one-hot, and the Gumbel-Softmax distribution converges to match the categorical distribution $Cat(\pi)$.


\subsection{Straight-through Gumbel-Softmax Estimator (STGS)}
The continuous relaxation of one-hot vectors is effective for tasks involving learning hidden representations and sequence modeling. However, in situations where sampling discrete values is necessary, (such as in a discrete action space for reinforcement learning or quantized compression, or searching for optimal architecture as in neural architecture search) we employ a discretization process of $S_\lambda$ using argmax. Despite this discretization step during the forward pass, we continue to leverage our continuous approximation in the backward pass by approximating the gradient of the relaxed continuous variable, and the STGS is given by:
\begin{equation}
    \nabla_{STGS}:=\frac{\partial\:f(S_\lambda)}{\partial\:S_\lambda}\:\frac{\partial\:S_\lambda}{\partial\:\phi}
\end{equation}

where $S_\lambda=softmax(\phi+G)$ with $\phi=log \theta$ and $G$ is the i.i.d Gumbel variable. 
More details can be found in the supplementary material.



\begin{figure*}[t]
  \centering
   \includegraphics[width=0.8\linewidth]{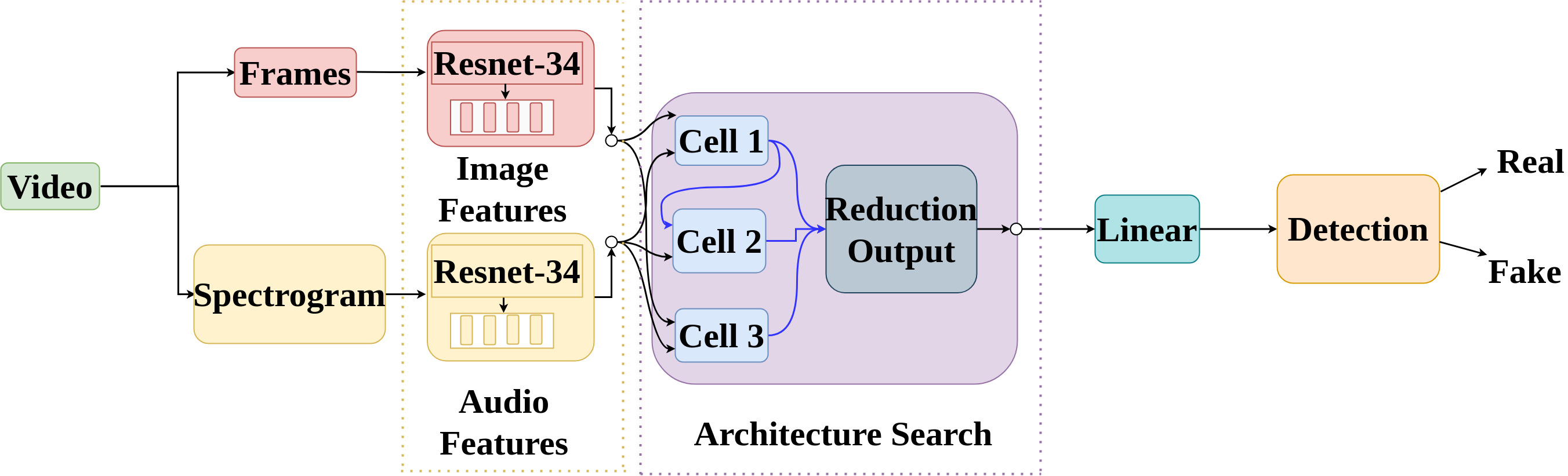}

   \caption{Block diagram indicating the multimodal fusion network proposed by STGS-BMNAS, which consists of two level searching scheme. In the first level we search for features from backbone network. Each cell accepts two inputs from its previous cells. In the second stage we search for optimal architecture searched using our proposed STGS over the cells through pool primitive operations and finally concatenate the cell outputs for prediction.}
   \label{fig:deepfake_arch}
\end{figure*}

\begin{figure*}[t]
  \centering
   \includegraphics[width=1 \linewidth, height=0.15 \linewidth]{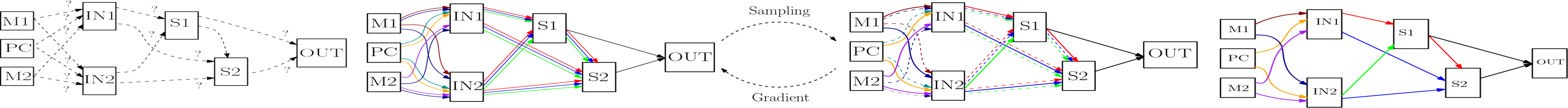}

   \caption{An overview of the conceptual visualization of our proposed STGS-BMNAS:
a) Initially, an acyclic graph is predefined, with cells receiving inputs from the backbone network.
b) During forward propagation for the first level search (indicated by colors), we utilize our proposed Gumbel Softmax to sample features from the backbone network. Subsequently, in the next stage, we also use the same Gumbel Softmax to sample an optimal architecture. During the backward pass, we employ Straight Through Estimator to simultaneously calculate gradients and network.
c) Finally, we obtain the final network using our proposed Straight Through Gumbel Softmax estimator.}
\label{fig:sam_grad}
{M1: Modality-1 in our case image, M2: Modality-2 in our case speech, PC: Previous cell, IN1: Input-1, IN2: Input-2, Intermediate nodes-S1:Step node-1 S2: Step node-2, OUT:Output node}
\end{figure*}

\subsection{Understanding our proposed STGS-BMNAS}
For better understanding of our STGS-BMNAS, we define four propositions.
\begin{enumerate}

\item  \textbf{Proposition-1:} For an arbitrary input $n$, the STGS operation is differentiable:
\begin{equation}
    \frac{\partial \tilde p_i}{\partial \theta_j}=\frac{\partial}{\partial \theta_j}Softmax \{\frac{log\ (\theta)_i+G^{(i)}}{\lambda}\}
\end{equation}
where $\tilde p_i$\>=sampled probability, $\theta_j$\>= logits corresponding to j, $P_i$\> = original probability of category i,$G^{(i)}$\> = Gumbel noise, $\lambda$\> = temperature parameter


\item  \textbf{Proposition-2:} As the temperature parameter $\lambda\rightarrow 0$, the STGS operation assumes categorical distribution.
\begin{equation}
\begin{split}
    &\lim_{\lambda\rightarrow 0}Softmax\left(\frac{\log \ \theta_i+G^{(i)}}{\lambda}\right)\\
    &\quad\rightarrow argmax\left(\frac{\log \ \theta_i+G^{(i)}}{\lambda}\right) = 
    \begin{cases}
        1 & \text{if}\ x > 0\\
        0 & \text{otherwise}
    \end{cases}
\end{split}
\end{equation}


This proposition ensures that as $\lambda$ decreases, the sampled probabilities becomes more concentrated, resembling one-hot vectors.

\item  \textbf{Proposition-3:} The temperature parameter $\lambda$ controls the entropy distribution of STGS.
\begin{equation}
    Softmax(\frac{log\ \theta_i+ G^{(i)}}{\lambda})=\frac{exp(log \ (\theta_i)+G^{(i)})}{\sum_{j=1}^{K}exp(\frac{log \ (\theta_i)+G^{(i)}}{\lambda})}
\end{equation}
as $\lambda<0$, the exponential term becomes more and more sensitive to differences in logits resulting in lower entropy and peakness.

\item  \textbf{Proposition-4:} For a K-dimensional probability vector $p=[p_1,p_2....p_k] $ with K categories and a number of sampling times M, STGS can generate a large number of distinct probabilities. Specifically, for each category i, STGS can sample M probabilities in range $[0,1]$ covering wide range of probabilities.


\end{enumerate}

\subsection{Single cue feature extraction}
In this work, similar to numerous other approaches to multimodal fusion \cite{perez2019mfas, yin2022bm}, we begin with the assumption of employing a pre-existing CNN feature extractor for each of the modalities involved. Practically, this entails initiating with a pre-trained CNN for each of the image and speech cues. We utilize a pre-trained ResNet-34 model trained on ImageNet for facial feature extraction, alongside MTCNN \cite{zhang2016joint} for face detection. Similarly for speech , we employ a pre-trained ResNet-34 model trained on Voxceleb data \cite{ding2020autospeech}. For each utterance a $300 \times 257$ dimensional spectrogram is extracted. We use the outputs generated from hidden layers or intermediate layers rather than final output layers. It captures the features and representations learned by the network at that specific layer, providing a more abstract and nuanced representation of the input data compared to the raw input or the final output.

\subsubsection{First level: Straight through Gumbel softmax relaxation over the cells}
In this level, we search for single modal features from the backbone networks. Formally for two cues image (I) and speech (S), let ${I^{i}}$ be the features extracted from backbone model and let ${S^{i}}$ be the features extracted from the second backbone model. Then we formulate the first level nodes in a sequence 
\begin{multline}
    \mathbb{R}=\{I^{(1)}, I^{(2)}...,I^{(N_I)},S^{(1)},S^{(2)}...,S^{(N_S)}... \\ Cell^{(1)}...,Cell^{(N)}\}
\end{multline}
Let $\mathbb{R}^u$, $\mathbb{R}^v$ be any two nodes from $\mathbb{R}$. Let $\alpha$ be the weight parameter connecting between $\mathbb{R}^{(u)}$, $\mathbb{R}^{(v) }$ then each edge is selected based on the unary operation. Let $\mathbb{O}^F$ be the set of candidate operations\\ 
 
$\mathbb{O}^F=\left\{ 
  \begin{array}{ c l }
  \centering
    Identity(x)=x & \quad \textrm{selecting an edge} \\
    Zero(x)=0                 & \quad \textrm{discarding an edge}
  \end{array}
\right.$
    
where each operation refers to a function o(.) to be applied on the $cell^{(u)}$ then by applying the gumbel softmax 
\begin{equation}
    \bar{o}(u,v)_\lambda= \sum_{o \in O} \frac{exp((\alpha_o^{(u,v)}+G^{(u)})/\lambda)}{\sum_{o^{'}\in O}exp((\alpha_o+G^{(v)})/\lambda)}o(x)
\end{equation}
where $\lambda$ is the temperature parameter that influences the entropy of the gumbel softmax distribution. Then the straight through gumbel softmax is given by the following equation:

\begin{equation}
    \nabla_{STGS}=\frac{\partial f(\bar{o}(u,v))}{\partial\bar{o}(u,v))}\;\frac{\partial\bar{o}(u,v)}{\partial\phi}
\end{equation}
where $\phi=log(\alpha)$.

A cell is densely connected and receives input from all its predecessors.
\begin{equation}
    o^v=\sum_{u<v} \bar{o}^{(u,v)}(o^{(i)})
\end{equation}

In the evaluation stage, since we want deterministic predictions, the probabilities obtained from Gumbel-softmax distribution can be directly used without the need for sampling or argmax operation as,

\begin{equation}
    (u,v)=\alpha_o(u,v)
\end{equation}

It is worth noting that, by directly using the probabilities, we obtain deterministic predictions without introducing randomness through sampling. The advantages of using of softmax probabilities rather than argmax during evaluation mode include probabilistic interpretations i,e., softmax provides probabilistic interpretations rather than single prediction (argmax) and softmax probabilities are robustness to noise when compared to argmax.
\begin{algorithm}[!h]
\caption{Optimization of Fusion Networks with Gumbel-Softmax Relaxation and Straight-Through Estimator}
\label{alg:fusion_optimization_gumbel_st}
\begin{algorithmic}[1]
    \STATE \textbf{Input:} Training data and validation data
    \STATE Initialize architecture parameters $\alpha, \beta, \gamma$ and model parameters $w$;
    \STATE Initialize discrete architecture based on $\alpha, \beta, \gamma$, set architecture\_best = architecture;
    \STATE Construct hypernet based on architecture\_best;
    \WHILE{$L$ not converged}
        \STATE Update $\omega$ on training set;
        \STATE Sample Gumbel noise $\mathbb{G} \sim \text{Gumbel}(0, 1)$;
        \STATE Compute logits for $\alpha$: $y_\alpha = \text{Softmax}\left(\frac{\log(\alpha) + \mathbb{G}}{\lambda}\right)$;
        \STATE Compute logits for $\beta$: $y_\beta = \text{Softmax}\left(\frac{\log(\beta) + \mathbb{G}}{\lambda}\right)$;
        \STATE Compute logits for $\gamma$: $y_\gamma = \text{Softmax}\left(\frac{\log(\gamma) + \mathbb{G}}{\lambda}\right)$;
        \STATE Sample architecture parameters $\hat{\alpha}$, $\hat{\beta}$, $\hat{\gamma}$ using Gumbel-Softmax reparameterization: $\hat{\alpha} = y_{\alpha_i}$, $\hat{\beta} = y_{\beta_i}$, $\hat{\gamma} = y_{\gamma_i}$;
        \STATE Update $(\alpha, \beta, \gamma)$ on validation set using $\hat{\alpha}$, $\hat{\beta}$, $\hat{\gamma}$;
        \STATE Derive upper level architecture based on $\alpha$, derive lower level architecture based on $\beta, \gamma$;
        \STATE Update hypernet based on architecture;
        \IF{higher validation accuracy is reached}
            \STATE Update architecture\_best using architecture;
        \ENDIF
        \STATE Compute gradients w.r.t. $\hat{\alpha}$, $\hat{\beta}$ and $\hat{\gamma}$ using straight-through estimator: $\nabla_{\hat{\alpha},\hat{\beta},\hat{\gamma}} L_{\text{val}} \approx \nabla_{\alpha,\beta,\gamma}, L_{\text{val}} \bigg|_{\alpha=\hat{\alpha},\beta=\hat{\beta},\gamma=\hat{\gamma}}$;
        
    \ENDWHILE
     
    \textbf{return} architecture\_best;
    
\end{algorithmic}
\end{algorithm}

\subsubsection{Second level: Weighted fusion}
Similar to \cite{yin2022bm} we use the same predefined candidate operations shown in the Table \ref{tab:second_cand}.

\begin{table*}[!t]
\centering

\begin{tabular}{c|c}
\hline
\textbf{Operation} & \textbf{Function}  \\ \hline
Zero(x,y) & \makecell{The Zero operation, eliminates an entire node,\\
effectively discarding its contribution}. \\ \hline
Sum(x, y): & \makecell{The DARTS  framework \cite{liu2018darts}, introduced , \\ employs a method to combine two features using summation. \\ \vbox{\begin{equation*}
    \centering
        Sum(x,y)=x+y
    \end{equation*}}}  \\ \hline
Attention$(x,y)$ & \makecell{ The Attention operation, as described in \cite{vaswani2017attention}, employs scaled dot-product attention,\\  where a query $x$ and key-value pairs $y$ are used. \\
\vbox{\begin{equation*}
        \centering
        Attention(x,y)=Softmax(\frac{xy^T}{\sqrt{C}} \times y)
    \end{equation*}}} \\ \hline
 LinearGLU$(x,y)$& \makecell{The LinearGLU operation combines two inputs $x,y$,\\ using a linear layer followed by the gated linear unit (GLU) activation \cite{yin2022bm}. \\
 \vbox{\begin{equation*}
       \centering
    LinearGLU(x,y)=xW_1\bigodot Sigmoid(yW_2)
   \end{equation*}}} \\ \hline
ConcatFC$(x,y)$ & \makecell{ The ConcatFC operation involves concatenating two inputs, \\
$(x,y)$ and and passing the concatenated vector through a fully connected (FC) layer with ReLU activation. \\ \vbox{\begin{equation*}
\centering
    ConcatFC(x,y)=ReLU((x,y).W+b)
\end{equation*}}}\\ \hline
\end{tabular}
\caption{Candidate operations used in the second level search}
\label{tab:second_cand}
\end{table*}

In the second level of STGS-BMNAS, searches for weighted fusion strategy within the cell. A cell is directed acyclic graph consists of ordered sequence of N nodes. Each nodes $n^i$ is a latent representation (eg., attention operation) and has directed edges $(u,v)$ associated with some operation $o^{(u,v)}$ that transforms $n^i$. 

\textbf{Weighted fusion strategy:} In this stage, the inner structure of $Cells^{(n)}$ is an ordered sequence of $\mathbb{C}_n$
\begin{equation}
    \mathbb{C}_n=\{I,S, N^{(1)},....N^{(M)}\}
\end{equation}
A cell consists of three nodes namely:
\begin{itemize}
   
\item  \textbf{Input node:} The input node $in_c^{(i)}$ receives the input from backbone network, and transforms into two other intermediate nodes.

\item  \textbf{Intermediate node:} The input node $in_c^{(i)}$ is transformed into intermediate nodes $c^{(j)}$, $c^{(k)}$ through straight through gumbel softmax weighted fusion over the candidate operations $\mathbb{O}^S$ 
\begin{multline}
\tiny
    \bar{o}(c^{(j)},c^{(k)})=\sum_{o^s \in \mathbb{O}^s} \frac{exp((\gamma^{(i)}+G^{(i)})/\lambda)}{\sum_{o' \in \mathbb{O}^s}exp((\gamma_{\bar{o}^s}^{(i)}+G^{(i)})/\lambda)} \times \\ w_i(f(c^{(j)},c^{k}))
\end{multline}
where $\gamma$ is the weight of candidate operations. In the evaluation stage since we want deterministic predictions, the probabilities obtained from Gumbel-softmax distribution can be directly used without the need for sampling or argmax operation as,

\begin{equation}
    o^{(i)}=\gamma_c^{(i)}
\end{equation}


The edge weights $(\beta)$ are also relaxed using straight-through gumbel softmax similar to the first level.





\item \textbf{Output node:} The output from all the transformational nodes are concatenated to form the output node.
The algorithm for two-level optimization with Straight-through Gumbel-softmax estimator is described in \ref{alg:fusion_optimization_gumbel_st}.
\end{itemize}

\subsection{Parameter Learning and Architecture Sampling}
To train the architecture in an end-to-end fashion, the weight parameters and architecture is simultaneously optimised with our proposed STGS. Analogous to the architecture search proposed \cite{yin2022bm}, the validation set performance is considered as a reward in our model, but training in differentiable manner through our proposed STGS. Denoting the loss as $\mathbb{L}_\omega$, the goal of the architecture search is to find a high performance  architecture with fewer model parameters i,e.,
\vspace{-0.3cm}

\begin{equation}
    \underset{\omega,\alpha,\gamma}{min} \mathbb{E}_{\mathbb{A}\sim p_{(\alpha,\gamma)_{\lambda}}}|\mathbb{L}_\omega(\mathbb{A}))|
\end{equation}

The main process of optimizing this objective is to minimise the the expected performance of the architecture sampled from 
\begin{equation*}
    p_{(\alpha,\gamma)_{\lambda}}(\mathbb{A})
\end{equation*}
The network architecture is first sampled from $p_{(\alpha,\beta,\gamma)}$ by varying the temperature parameter $\lambda$ afterwards the loss on the training is set is calculated in forward propagation and relying on this loss, the parameters ($\alpha,\gamma$) and network parameter $\omega$ is obtained to modify these parameters better. During back propagation, our model is trained in an end-to-end manner via straight through estimator. In the end an optimal architecture with  fewer model parameters is obtained by choosing optimal values for temperature parameter $\lambda$ and corresponding architecture parameters $\alpha,\gamma$. A conceptual visualisation of our STGS is shown in Figure \ref{fig:sam_grad}.

\section{Experiments and Results}
\subsection{Datasets}
\begin{enumerate}
    \item \textbf{FakeAVCeleb:} We run our experiments on the audio-visual FakeAVCeleb dataset \cite{khalid2021fakeavceleb}, The dataset described was created in 2022 specifically for the task of deepfake detection. It consists of a compilation of 500 YouTube videos featuring 500 celebrities from various ethnic backgrounds who are prominent figures in music, movies, sports, politics, and other fields. This database has 19,500 Fake Videos and 500 Real videos.
    \item \textbf{SWAN-DF:} The SWAN-DF dataset \cite{khalid2021fakeavceleb} represents the first high-fidelity collection of realistic audio-visual deepfakes that has been made publicly accessible. It is built upon the SWAN (Shared Wireless Access Network) database, which originally consisted of actual high-definition movies captured using iPhones and iPad Pro. This database has 24,000 fake videos and 2,800 real videos.
\end{enumerate}
 
\subsection{Performance evaluation protocol:} 
We perform two different types of evaluation. i) we mix two databases and evaluate the performance on the test set. ii) we check for ability of our model to generalise to different dataset. We achieve these by perform architecture search on FakeAVCeleb dataset and evaluate the model on  FakeAVCeleb and unseen SWAN-DF dataset and vice versa. Since both databases are biased towards fake videos than the real videos. To mitigate this bias we apply 36 different augmentation for train and validation split of both database. After applying augmentation we have 50,742 training videos, 10,718 validation videos and 8,963 test videos. Further details about the dataset split and different augmentations applied is given in the supplementary material.

\begin{table*}[!t]
\centering
\footnotesize
\begin{tabular}{c|c|c|c|c|c}
\hline
\textbf{Method} & \textbf{AUC}(\%) &\textbf{ACC}(\%) &\textbf{Params(M)}  &\textbf{\makecell{GPU \\ days}}  &\textbf{Search}  \\ \hline
Voice-face \cite{cheng2023voice}&82  & 86 &174  & - &Gradient \\ 
\makecell{Audio-visual \\ anamoly detection \cite{feng2023self}}&93  & - &41  &- &Gradient \\ 
\makecell{Not made \\ for each other \cite{chugh2020not}} & 81 & 84.56 &122  &- &Gradient  \\
 ID-Reveal \cite{cozzolino2021id}& 78 & 80.1 &7.3  &- &Gradient \\ 
MultimodalTrace \cite{raza2023multimodaltrace} & 84 & 91.26 &15  &-  &Gradient\\ 
 Ensemble-learning \cite{hashmi2022multimodal}& 84 & 86 &12  &- & Gradient\\ 
POI-AV \cite{sung2023hearing} & 93.9 & 90.9&-  &- &- \\ 
BM-NAS \cite{yin2022bm} &92.26  & 91.4 &0.62  & 4&Gradient \\ 
     \textbf{STGS-BMNAS (Ours)}& \textbf{94.4} & \textbf{95.5} &\textbf{0.26}  & 2&\textbf{\makecell{Straight\\ through estimator}} \\ \hline
 
\end{tabular}
\caption{Comparison of our proposed STGS-BMNAS with SOTA approaches tested on our test data}
\label{tab:SOTA_cmp}
\end{table*}


\begin{figure}[!h]
    \centering
    \begin{minipage}{.15\textwidth}
        \includegraphics[width=\linewidth]{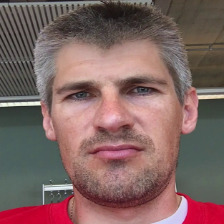}
    \end{minipage}
    \begin{minipage}{.15\textwidth}
        \includegraphics[width=\linewidth]{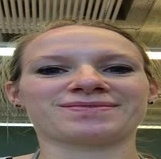}
    \end{minipage}
    \begin{minipage}{.15\textwidth}
        \includegraphics[width=\linewidth]{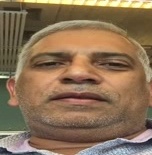}
    \end{minipage}
    
    \caption*{(a)}
    \label{fig:first_row}
    \\
    \begin{minipage}{.15\textwidth}
        \includegraphics[width=\linewidth]{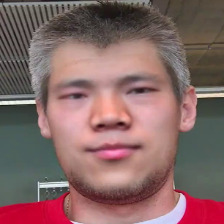}
    \end{minipage}
    \begin{minipage}{.15\textwidth}
        \includegraphics[width=\linewidth]{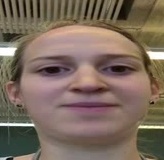}
    \end{minipage}
    \begin{minipage}{.15\textwidth}
        \includegraphics[width=\linewidth]{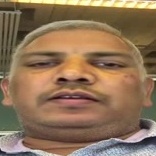}
    \end{minipage}
    \caption*{(b)}
    \caption{Sample images of real and fake a) Real samples of FakeAVCeleb and SWAN-DF b) Fake samples from FakeAVCeleb and SWAN-DF}
    \label{fig:data_samples}
\end{figure}
\begin{figure}[!h]
  \centering
   \includegraphics[width=0.8\linewidth]{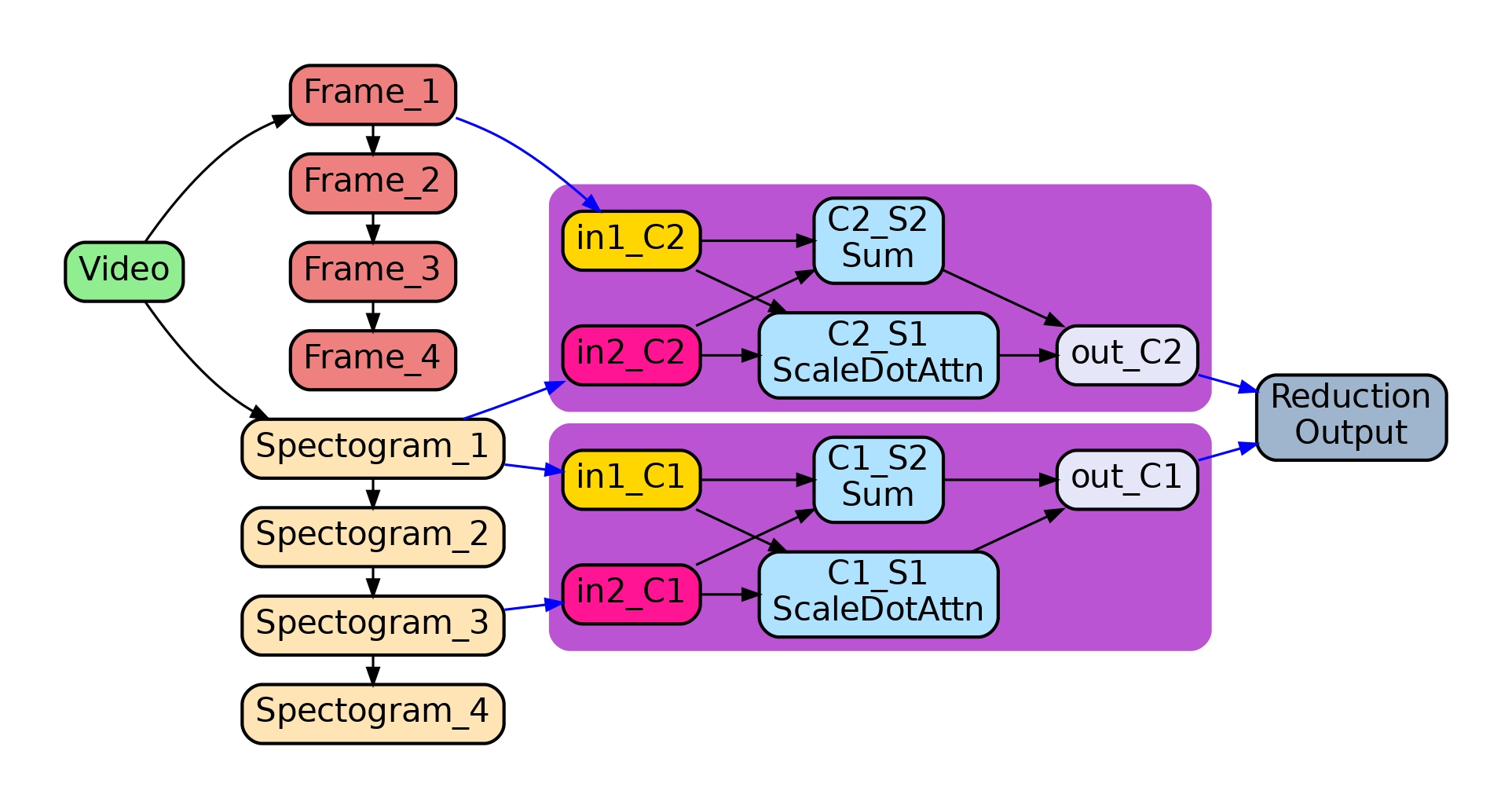}

   \caption{Optimal architecture obtained with temperature $\lambda=10$ and sampling M=15 for second type of evaluation protocol.}
   \label{fig:optimal_gumbel_softmax}
\end{figure}

\subsection{Architecture search on our dataset}

In our experiments, we consider two sets of operations. The first set involves selecting or discarding edges from the pool of operations denoted as $\mathbb{O}^F$. In the second stage, we consider operations from another pool denoted as $\mathbb{O}^S$. We vary the sampling times M for different values of the temperature parameter $\lambda$, and we choose $\lambda=10$ and sampling times of M=15 for rich search space and for lower training loss (see Training loss and validation loss graph for different sampling times M in supplementary material). The algorithm terminates when the choice of operations in the neural cell converges for both learnable parameters $\alpha$ and $\gamma$ which is empirically measured as:
\begin{equation}
    E(\alpha)=-\sum_{u,v}\sum_{o \in \mathbb{O}^F} \alpha^o_{(u,v)}\:log(\alpha^o_{uv})
    \label{eq:entropy_alpha}
\end{equation}
\begin{equation}
    E(\gamma)=-\sum_{u,v}\sum_{o \in \mathbb{O}^S} \alpha^o_{(u,v)}\:log(\gamma^o_{uv})
    \label{eq:entropy_gamma}
\end{equation}

The architecture search was conducted on V100 Tesla GPUs, each equipped with a memory size of 16GB. The training configurations included an architecture learning rate of 0.003, architecture weight decay of 0.001, momentum set to 0.9, and weight decay of 0.003. The optimizer employed was Adam, with a maximum learning rate of 0.003 and a minimum learning rate of 0.006. The batch size set at 8, and the training was implemented using the PyTorch framework. If the entropy remains constant for dozens of epochs according to equations \ref{eq:entropy_alpha} ,\ref{eq:entropy_gamma}, we conclude that the architecture search is converged.


\subsection{Architecture evaluation on our dataset}
To evaluate the searched architecture, we train the optimal architecture obtained from the search for 100 epochs with a batch size of 64 and report its performance on a test set. The architecture evaluation is carried out with similar configurations used in training but with a batch size of 16, dropout probability of 0.2, maximum learning rate of 0.003, and minimum learning rate of 0.006.

\begin{table}[!h]
\footnotesize
\begin{tabular}{|cc|cccc}

\hline
\multicolumn{2}{c|}{\multirow{2}{*}{ \textbf{Temperature ($\lambda$)} }}    & \multicolumn{4}{c}{\textbf{Sampling times (M)}}                                                    \\ \cline{3-6}
\multicolumn{2}{c|}{}                     & \multicolumn{1}{c|}{\textbf{5}} & \multicolumn{1}{c|}{\textbf{10}} & \multicolumn{1}{c|}{\textbf{15}} &\textbf{20}  \\ \hline
\multicolumn{1}{c|}{\multirow{2}{*}{$\lambda=5$}} & AUC  & \multicolumn{1}{c|}{86.7} & \multicolumn{1}{c|}{87.3} & \multicolumn{1}{c|}{89.2} & 91.4 \\  
\multicolumn{1}{c|}{}                  & \makecell{Model \\ parameters} & \multicolumn{1}{c|}{651776} & \multicolumn{1}{c|}{683776} & \multicolumn{1}{c|}{534400} & 404096 \\ \hline
\multicolumn{1}{c|}{\multirow{2}{*}{$\lambda=10$}} &AUC  & \multicolumn{1}{c|}{92} & \multicolumn{1}{c|}{94} & \multicolumn{1}{c|}{92} &90  \\  
\multicolumn{1}{c|}{}                  & \makecell{Model \\ parameters} & \multicolumn{1}{c|}{356096} & \multicolumn{1}{c|}{257920} & \multicolumn{1}{c|}{255872} & 223187 \\ \hline
\multicolumn{1}{c|}{\multirow{2}{*}{$\lambda=15$}} & AUC & \multicolumn{1}{c|}{94.5} & \multicolumn{1}{c|}{95.1} & \multicolumn{1}{c|}{95.6} &96.3  \\  
\multicolumn{1}{c|}{}                  & \makecell{Model \\ parameters} & \multicolumn{1}{c|}{2165674} & \multicolumn{1}{c|}{201964} & \multicolumn{1}{c|}{196186} & 185726 \\ \hline
\multicolumn{1}{c|}{\multirow{2}{*}{$\lambda=20$}} & AUC & \multicolumn{1}{c|}{96.25} & \multicolumn{1}{c|}{96.4} & \multicolumn{1}{c|}{97.1} &97.5  \\  
\multicolumn{1}{c|}{}                  & \makecell{Model \\ parameters} & \multicolumn{1}{c|}{183264} & \multicolumn{1}{c|}{177624} & \multicolumn{1}{c|}{170125} & 162517 \\ \hline
\end{tabular}
\caption{Evaluation of searched architecture with different temperature values and with varying sampling values}
\label{tab:ablation_temp_sample}
\end{table}

\subsection{Comparative performance analysis}
We use two metrics for deepfake detection named Area Under Curve (AUC) and standard classification accuracy (ACC) similar to the studies of \cite{sung2023hearing}. 


\textbf{State-of-the-art approaches:} We consider publicly available implementations for fair comparison. We consider only audio-visual deepfake detection methods as our proposed method requires both modalities for classification. The SOTA approaches considered for comparison are: Not made for each other \cite{chugh2020not}, Voice-Face \cite{cheng2023voice}, Audio-Visual Anamoly detection \cite{feng2023self}, ID-Reveal \cite{cozzolino2021id}, Multimodaltrace \cite{raza2023multimodaltrace}, Ensemble learning \cite{hashmi2022multimodal} and POI-AV \cite{sung2023hearing}.

\textbf{Training:} For fair comparison we train all the methods with our dataset and follow the pre-processing steps mentioned in the methods and we also ensure that training, evaluation, and test data do not overlap.

Table \ref{tab:SOTA_cmp} shows the comparative analysis of our proposed STGS-BMNAS with SOTA methods by mixing both databases. Our method significantly outperforms recently published POI-AV \cite{sung2023hearing}, Multimodaltrace \cite{raza2023multimodaltrace}, ID-Reveal \cite{cozzolino2021id} both in terms of ACC and AUC metrics with very fewer model parameters. However, BM-NAS \cite{yin2022bm} achieved similar performance in terms of model parameters, but unlike our method, the training graph does not converge (Figure is given in the Supplementary material). The ROC curves for constant temperature parameter $\lambda=10$ and 
M=15 sampling times has the highest AUC value when compared to other sampling numbers(Figure is given in the Supplementary material).

\textbf{Generalization of the model to the dataset:}
The results shown in the Table \ref{tab:gen_data} demonstrate how effectively the proposed model generalizes to unseen data. Our model showcases decent performance when tested to unseen data, underscoring its robust generalizability for such scenarios. However for trained and tested on same database our proposed model shows superior performance. The sample architectures for this case are shown in the supplementary material.

\begin{table}[!h]
\begin{tabular}{|c|cccc|}
\hline
\multirow{3}{*}{\makecell{\textbf{Trained on} \\$\downarrow$}} & \multicolumn{4}{c|}{\makecell{\textbf{Tested on} \\ $\downarrow$}}                                                    \\ \cline{2-5} 
                  & \multicolumn{2}{c|}{FakeAVCeleb}                         & \multicolumn{2}{c|}{SWAN-DF}    \\ \cline{2-5} 
                  & \multicolumn{1}{c|}{AUC} & \multicolumn{1}{c|}{ACC} & \multicolumn{1}{c|}{AUC} &ACC \\ \hline
 FakeAVCeleb                 & \multicolumn{1}{c|}{\textbf{92.7}} & \multicolumn{1}{c|}{\textbf{91.8}} & \multicolumn{1}{c|}{85.6} &  84.7\\ \hline
    SWAN-DF              & \multicolumn{1}{c|}{84.8} & \multicolumn{1}{c|}{83.2} & \multicolumn{1}{c|}{\textbf{93.1}} & \textbf{92.8} \\ \hline
\end{tabular}
\caption{Generalisation of our proposed model to the seen and unseen data}
\label{tab:gen_data}
\end{table}
 

\subsection{Ablation study}
An ablation study is conducted for different values of temperature and with varying sampling time and a search is conducted. In the evaluation stage with the obtained architecture we calculate the model parameters and AUC values. Table \ref{tab:ablation_temp_sample} shows the ablation study results. From the table we see that if the temperature value and sampling value is increased we will get better AUC value and very fewer model parameters. This is in accordance to the Proposition 3, i.e., as the temperature values are increased more optimal architecture can be found with larger M but at the cost of more GPU days. Hence we choose an optimal temperature i.e., $\lambda=10$ and sampling M=15. The sample architectures obtained with different temperature and sampling parameters are as shown in supplementary material.

\section{Conclusion}
In this paper we propose a novel neural network architecture called STGS-BMNAS for deepfake detection. This architecture leverages a two-level schema to learn an optimal architecture by exploring architectural possibilities. It first identifies pertinent features from backbone networks and then conducts a search for the optimal architecture using both image and speech features. The network incorporates a weighted fusion operation to combine information from multiple sources. The proposed architecture achieves high classification performance while maintaining a low number of model parameters.



\end{document}